# Machine learning and glioblastoma: treatment response monitoring biomarkers in 2021

Thomas C. Booth,[1,2*] Bernice Akpinar[1], Andrei Roman[3,4], Haris Shuaib[5,6], Aysha Luis[1,7], Alysha Chelliah[1], Ayisha Al Busaidi[2], Ayesha Mirchandani[8], Burcu Alparslan[2,9], Nina Mansoor[10], Keyoumars Ashkan[11], Sebastien Ourselin[1] and Marc Modat[1]

[1] School of Biomedical Engineering & Imaging Sciences, King's College London, St. Thomas' Hospital, London SE1 7EH, UK

[2] Department of Neuroradiology, King's College Hospital NHS Foundation Trust, London SE5 9RS, UK

[3] Department of Radiology, Guy's & St. Thomas' NHS Foundation Trust, London SE1 7EH, UK

[4] The Oncology Institute "Prof. Dr. Ion Chiricuţă" Cluj-Napoca, Strada Republicii 34-36, Cluj-Napoca 400015, Romania

[5] Department of Medical Physics, Guy's & St. Thomas' NHS Foundation Trust, London SE1 7EH, UK

[6] Institute of Psychiatry, Psychology & Neuroscience, King's College London, London, SE5 8AF, UK

[7] Lysholm Department of Neuroradiology, National Hospital for Neurology and Neurosurgery, Queen Square, London, UK

[8] Cambridge University Hospitals NHS Foundation Trust, Hills Road, Cambridge, CB2 0QQ, UK

[9] Department of Radiology, Kocaeli Universityİzmit, Kocaeli, Turkey

[10] Department of Radiology, King's College Hospital NHS Foundation Trust, London SE5 9RS, UK

[11] Department of Neurosurgery, King's College Hospital NHS Foundation Trust, London SE5 9RS, UK

*Corresponding author Thomas C. Booth : tombooth@doctors.org.uk

ORCID ID: https://orcid.org/0000-0003-0984-3998

**Abstract.** The aim of the systematic review was to assess recently published studies on diagnostic test accuracy of glioblastoma treatment response monitoring biomarkers in adults, developed through machine learning (ML). Articles published 09/2018–09/2020 were searched for using MEDLINE, EMBASE, and the Cochrane Register. Included study participants were adult patients with high grade glioma who had undergone standard treatment (maximal resection, radiotherapy with concomitant and adjuvant temozolomide) and subsequently underwent follow-up imaging to determine treatment response status (specifically, distinguishing progression/recurrence from progression/recurrence mimics - the target condition). Risk of bias and applicability was assessed with QADAS 2 methodology. Contingency tables were created for hold-out test sets and recall, specificity, precision, F1-score, balanced accuracy calculated. Fifteen studies were included with 1038 patients in training sets and 233 in test sets. To determine whether there was progression or a mimic, the reference standard combination of follow-up imaging and histopathology at re-operation was applied in 67% (10/15) of studies. External hold-out test sets were used in 27% (4/15) to give ranges of diagnostic accuracy measures: recall = 0.70-1.00; specificity = 0.67-0.90; precision = 0.78-0.88; F1 score = 0.74-0.94; balanced accuracy = 0.74-0.83; AUC = 0.80-0.85. The small numbers of patient included in studies, the high risk of bias and concerns of applicability in the study designs (particularly in relation to the reference standard and patient selection due to confounding), and the low level of evidence, suggest that limited conclusions can be drawn from the data. There is likely good diagnostic performance of machine learning models that use MRI features to distinguish between progression and mimics. The diagnostic performance of ML using implicit features did not appear to be superior to ML using explicit features. There are a range of ML-based solutions poised to become treatment response monitoring biomarkers for glioblastoma. To achieve this, the development and validation of ML models require large, well-annotated datasets where the potential for confounding in the study design has been carefully considered. Therefore, multidisciplinary efforts and multicentre collaborations are necessary.

# 1. Introduction

Glioblastoma, the commonest primary malignant brain tumour has a median overall survival of 14.6 months despite standard of care treatment which consists of maximal debulking surgery and radiotherapy, with concomitant and adjuvant temozolomide [1]. After treatment, monitoring biomarkers are required to detect any change in the extent of disease or provide evidence of treatment response [2]. Magnetic resonance imaging (MRI ) is particularly useful as it is non-invasive and captures the entire tumour volume and adjacent tissues and has been incorporated into recommendations for determining treatment response in trials [3,4]. However, false-positive progressive disease (pseudoprogression) may occur within 6 months of chemoradiotherapy, typically determined by changes in contrast enhancement on $T_1$-weighted MRI images, representing non-specific blood brain barrier disruption [5,6]. Because true progression is associated with worse clinical outcomes, a monitoring biomarker that reliably differentiates pseudoprogression and true progression would allow an early change in treatment strategy with termination of ineffective treatment and the option of implementing second-line therapies [7]. Because pseudoprogression is common occurring in approximately 10-30% of cases [8,9], the neuro-oncologist is commonly presented with the difficult decision as to whether to continue adjuvant temozolomide or not. Distinguishing pseudoprogression and true progression has been an area of research with significant potential clinical impact for more than a decade.

Pseudoprogression is an early-delayed treatment effect, in contrast to the late-delayed radiation effect (or radiation necrosis) [10]. Whereas pseudoprogression occurs during or within 6 months of chemoradiotherapy, radiation necrosis occurs after this period, but with an incidence that is an order of magnitude less than the earlier pseudoprogression [11]. Nonetheless, in the same way that it would be beneficial to have a monitoring biomarker that discriminates true progression from pseudoprogression, an imaging technique that discriminates true progression from radiation necrosis would also be beneficial to allow the neuro-oncologist to know whether to implement second-line therapies or not.

Multiple studies have attempted to develop monitoring biomarkers to determine treatment response. Many incorporate machine learning (ML) as a central pillar of the process. A review of studies up to 2018 showed that those taking advantage of enhanced computational processing power to build neuro-oncology monitoring biomarker models, for example using convolutional neural networks (CNNs), have yet to show benefit compared to ML techniques using explicit feature engineering and less computationally expensive classifiers, for example using multivariate logistic regression [11]. It is also notable that studies applying ML to build neuro-oncology monitoring biomarker models have yet to show overall advantage over those using traditional statistical methods. The discipline of applying radiomic studies to neuro-oncology is expanding and evolving rapidly thereby motivating the need to appraise the latest evidence as the findings from 2018 may have been superseded.

Building on previous work [11,12], the aim of the study is to systematically review recently published studies on diagnostic accuracy of treatment response monitoring biomarkers developed through ML for patients with glioblastoma.

## 2 Methods

This systematic review was prepared according to the Preferred Reporting Items for Systematic Reviews and Meta-Analysis: Diagnostic Test Accuracy (PRISMA-DTA) [13] and informed by Cochrane review methodology with emphasis on developing criteria for including studies [14], searching for studies [15], and assessing methodological quality [16].

## 2.1 Search strategy and selection criteria

MEDLINE, EMBASE and the Cochrane Register were searched using a wide variety of search terms, both text words and database subject headings to describe each concept for original research articles published between September 2018 and September 2020 (Supplementary Table S1). The search was not limited by language. A search for pre-prints and other non-peer reviewed material was not carried out. Included study participants were adult patients with high grade glioma who had undergone standard treatment (maximal resection, radiotherapy with concomitant temozolomide and adjuvant temozolomide) and subsequently underwent follow up imaging in order to determine treatment response status (specifically, distinguishing progression/recurrence from progression/recurrence mimics - the target condition).

Studies were excluded if they were focused on paediatrics, pseudoresponse (bevacuzimab-related response mimic), had no ML algorithm used in feature extraction, selection, or classification/regression. The treatment response outcome for the index test was determined by the ML model. The treatment response outcome for the reference standard was determined either by subsequent follow-up imaging or histopathology at re-operation or a combination of both. The reference list of each acquired article for relevant studies was checked manually.

## 2.2 Data extraction and risk of bias assessment

Risk of bias and concerns regarding applicability for each study were assessed using QADAS 2 methodology [17] and modified proformas. Data was extracted to determine whether the gliomas analysed were glioblastomas, anaplastic astrocytomas, anaplastic oligodendrogliomas or a combination of all (high-grade glioma); the ML technique used for the index test including cross validation techniques; information on training and hold-out test sets; the reference standard employed; MRI sequence type and non-imaging features used for analysis.

Reference standard follow-up imaging protocols were extracted and assessed for appropriateness. One assessment was the handling of confounding factors of second-line drug therapy, discontinuation of temozolomide and steroid use. The appropriateness of the treatment response term (target condition) used in the published study was also assessed. Evidence shows that contrast enhancing lesions enlarging for the first time due to pseudoprogression typically occur 0 - 6 months after chemoradiation and that contrast enhancing lesions enlarging for the first time due to radiation necrosis typically occur beyond 6 months after chemoradiation. If "post treatment related effects" (PTRE) was used as a treatment response term this incorporates both pseudoprogression and radiation necrosis [18,19]. Any deviation in the use of the three terms defined here was recorded. Information on the duration of follow-up imaging after contrast enhancing lesions enlarged was also extracted and assessed. Optimal strategies to determine any PTRE or progression/recurrence included assigning the baseline scan after chemoradiation, excluding $T_2$-w image enlargement [20] and allowing 6-month follow up from the time of contrast enhancement with two follow up scans to mitigate capturing the upslope of PTRE again over a short interval [21,22].

## 2.3 Data synthesis and statistical analysis

Using the published study data, 2 x 2 contingency tables were created for hold-out test sets and the principal diagnostic accuracy measures of recall (sensitivity), specificity, precision (positive predictive value), F1-score and balanced accuracy calculated. Reported area under the receiver operating characteristic curve (AUC) values and confidence intervals from the published study were collated. In studies where principal diagnostic accuracy measures and the available published study raw data did not match, the discrepancy was highlighted - a calculation was made for all diagnostic accuracy measures based on those diagnostic accuracy measures that had been published, rather than available raw data. If there were both internal and external hold-out test sets, only the external test set principal diagnostic accuracy measures were calculated. In cases where there was no hold-out test set, "no test set" was recorded and a brief summary of the training set principal diagnostic accuracy measures was recorded. The unit of assessment was per-patient.

**2.4 Subgroup analysis: prognostic biomarkers to predict subsequent treatment response**

Prognostic imaging biomarkers applied to glioblastoma typically predict overall survival from baseline images. However, we included a subgroup of studies whose method was to apply ML models to baseline images to serve as prognostic biomarkers to predict subsequent treatment response. The studies were analysed using the same methodology.

# 3 Results

### 3.1 Characteristics of included studies and bias assessment

Figure 1 shows that overall, 2017 citations met the search criteria, and the full text of 43 potentially eligible articles was scrutinized. Fifteen studies, of which 13 were retrospective, from September 2018 to September 2020 (including online first article publication prior to September 2018) were included. The total number of patients in training sets was 1038 and in test sets 233. The characteristics of 12 studies are shown in Table 1 and the characteristics of 3 studies applying ML models to baseline images (or genomic alterations) to serve as prognostic biomarkers to predict subsequent treatment response are shown in Table 2.

The risk of bias assessment was performed for each study and summarised (Supplementary Figure S1). All or most of the studies were in the highest category for risk of bias relating to the reference standard (15/15, 100%) and patient selection (13/15, 87%). A third to a quarter of studies relating to the index test (5/15, 33%) and flow and timing (4/15, 27%) were in the highest category for risk of bias or the risk was unclear, respectively. In terms of concerns regarding applicability, the results largely mirrored the risk of bias.

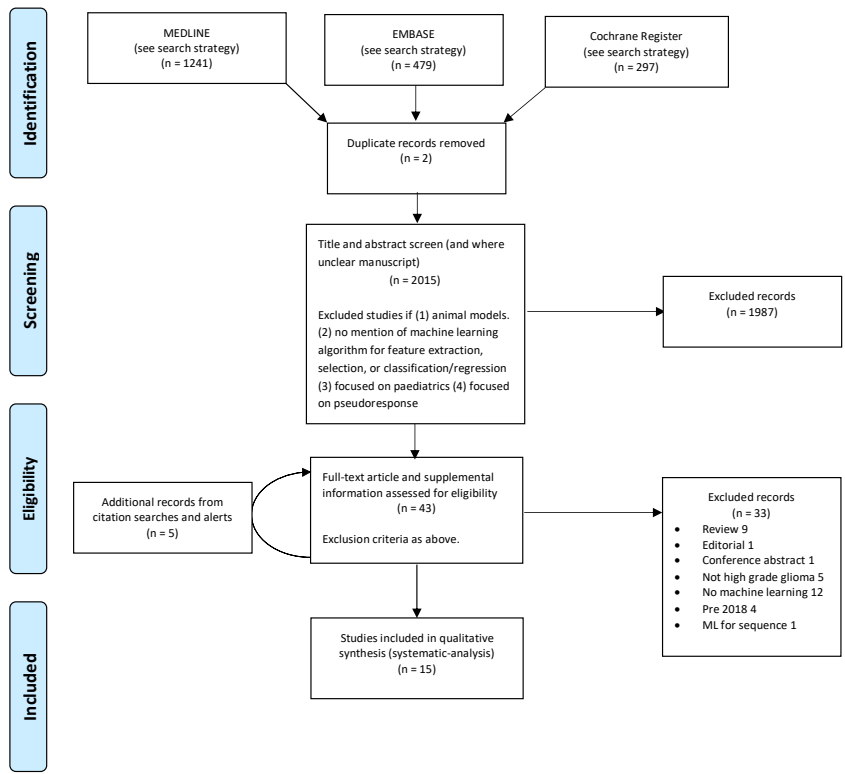

**Figure 1.** Flow diagram of search strategy.

**Table 1.** Studies applying machine learning to the development of high-grade glioma monitoring biomarkers

| Author(s) | Prediction | Reference | Dataset | Method | Features selected | Test set performance |
|---|---|---|---|---|---|---|
| *Kim JY et al., 2019[27] | Early True progression vs. early pseudoprogression | Mixture of histopathology & imaging follow-up | training = 61 testing = 34 $T_1$ C, FLAIR, DWI, DSC | Retrospective 2 centres: 1 train & 1 external test set. LASSO feature selection with 10-fold CV Linear generalized model | First-order, (volume/shape), Second-order (texture), wavelet. ADC & CBV parameters included. | Recall 0.71 Specificity 0.90 Precision 0.83 F1 0.77 BA 0.81 AUC 0.85 (CI 0.71–0.99) |
| Kim JY et al., 2019[28] | Early True progression vs. early pseudoprogression | Mixture of histopathology & imaging follow-up | training = 59 testing = 24 $T_1$ C, FLAIR, DTI, DSC | Retrospective 1 centre LASSO feature selection with 10-fold CV Linear generalized model | First-order, Second-order (texture), wavelet. FA & CBV parameters included. | Recall 0.80 Specificity 0.63 Precision 0.36 F1 0.50 BA 0.72 AUC 0.67 (0.40–0.94) |
| Bacchi S et al., 2019[29] | True progression vs. PTRE (HGG) | histopathology for progression & imaging follow-up for pseudoprogression | training = 44 testing = 11 $T_1$ C, FLAIR, DWI | Retrospective 1 centre 3D CNN & 5-fold CV | CNN. FLAIR & DWI parameters | Recall 1.00 Specificity 0.60 Precision 0.75 F1 0.86 BA 0.80 AUC 0.80 |
| Elshafeey N et al., 2019[30] | True progression vs. PTRE** | histopathology | training = 98 testing = 7 DSC, DCE | Retrospective 3 centres MRMR feature selection. 1 test. (i) decision tree algorithm C5.0 (ii) SVM including LOO & 10-fold CV | $K_{trans}$ & CBV parameters | Insufficient published data to determine diagnostic performance (CV training results available recall 0.91; specificity 0.88) |
| Verma G et al., 2019[x31] | True progression vs. Pseudoprogression | Mixture of histopathology & imaging follow-up | training = 27 3D-EPSI | Retrospective 1 centre Multivariate logistic regression LOOCV | Cho/NAA & Cho/Cr | No test set (CV training results available recall 0.94; specificity 0.87) |
| Ismail M et al., 2018[32] | True progression vs. Pseudoprogression | Mixture of histopathology & imaging follow-up | training = 59 testing = 46 $T_1$ C, $T_2$/FLAIR | Retrospective 2 centres: 1 train & 1 external test set. SVM & 4-fold CV | Global & curvature shape | Recall 1.00 Specificity 0.67 Precision 0.88 F1 0.94 BA 0.83 |
| *Bani-Sadr A et al., 2019[33] | True progression vs. Pseudoprogression | Mixture of histopathology & imaging follow-up | training = 52 testing = 24 $T_1$ C, FLAIR | Retrospective 1 centre Random Forest. | Second-order features +/- | Recall 0.94 (0.71-1.00) Specificity 0.38 (0.09-0.76) Precision 0.36 |
| | | | | | MGMT promoter status | F1 0.84<br>BA 0.66<br>AUC 0.77 |
| | | | | | | + non-MRI:<br>Recall 0.80 (0.56-0.94)<br>Specificity 0.75 (0.19-0.99)<br>Precision 0.86<br>F1 0.83<br>BA 0.74<br>AUC 0.85 |
| Gao XY et al., 2020[34] | True progression vs PTRE | Mixture of histopathology & imaging follow-up | training = 34<br>testing = 15<br>$T_1$ C, FLAIR | Retrospective 2 centres SVM & 5-fold CV | $T_1$ C, FLAIR subtraction map parameters | Recall 1.00<br>Specificity 0.90<br>Precision 0.83<br>F1 0.91<br>BA 0.95<br>AUC 0.94 (0.78–1.00) |
| Jang B-S et al., 2018[35] | True progression vs. Pseudoprogression | Mixture of histopathology & imaging follow-up | training = 59 & testing = 19<br>$T_1$ C & clinical features & IDH/ MGMT promoter status | Retrospective 2 centres 1 train & 1 external test set.<br>CNN LSTM & 10-fold CV (compared to Random Forest) | CNN $T_1$ C parameters +/- Age, Gender, MGMT status, IDH mutation, radiotherapy dose & fractions, follow up interval | Recall 0.64<br>Specificity 0.50<br>Precision 0.64<br>F1 0.63<br>BA 0.57<br>AUC 0.69 |
| | | | | | | + non-MRI:<br>Recall 0.72<br>Specificity 0.75<br>Precision 0.80<br>F1 0.76<br>BA 0.74<br>AUC 0.83 |
| Li M et al., 2020[36] | True progression vs PTRE** | Imaging follow-up | training = 84<br>DTI | Retrospective 1 centre DCGAN & AlexNet CNN with SVM including 5 & 10 & 20-fold CV (compared to DCGAN, VGG, ResNet, and DenseNet) | CNN DTI | No test set<br><br>(CV training results available recall 0.98; specificity 0.88; AUC 0.95) |
| Akbari H et al., 2020[37] | True progression vs. Pseudoprogression | histopathology | training = 40<br>testing = 23<br>testing = 20<br>$T_1$ C, $T_2$/ FLAIR, DTI, DSC, DCE | Retrospective 2 centres. 1 train & test. 1 external test set. imagenet_vgg_f CNN SVM & LOOCV | First-order, second-order (texture). CBV, PH, TR, $T_1$ C, $T_2$/FLAIR parameters included. | Recall 0.70<br>Specificity 0.80<br>Precision 0.78<br>F1 0.74<br>BA 0.75<br>AUC 0.80 |
| Li X et al., 2018[38] | Early True progression vs. early pseudoprogression (HGG) | Mixture of histopathology & imaging follow-up | training = 362<br>$T_1$ C, $T_2$, multi-voxel & single | Gabor dictionary & sparse representation classifier (SRC) | sparse representations | No test set<br><br>(CV training results available recall 0.97; specificity 0.83) |

| | | | voxel 1H-MRS, ASL | | | |
|---|---|---|---|---|---|---|

PTRE = post treatment related effects
HGG = high grade glioma

Magnetic resonance imaging sequences: $T_1C$ = post contrast $T_1$-weighted; $T_2$ = $T_2$-weighted; FLAIR = fluid-attenuated inversion recovery; DWI = diffusion-weighted imaging; DCE = dynamic contrast-enhanced; DSC = dynamic susceptibility-weighted; DTI = diffusor tensor imaging; ASL = arterial spin labelling.
Magnetic resonance imaging parameters: CBV = cerebral blood volume; PH = peak height; ADC = apparent diffusion coefficient; FA = fractional anisotropy; TR = trace (DTI); $K_{trans}$ = volume transfer constant.
1H-MRS = 1H-magnetic resonance spectroscopy; 3D-EPSI = 3D echo planar spectroscopic imaging.
Magnetic resonance spectroscopy parameters: Cho = choline; NAA = N-acetyl aspartate; Cr = creatine.

Molecular markers: MGMT = $O^6$-methylguanine-DNA methyltransferase; IDH = isocitrate dehydrogenase.

Machine learning techniques: CNN = convolutional neural network; SVM = support vector machine; LASSO = least absolute shrinkage and selection operator; mRMR = minimum redundancy and maximum relevance; CV = cross validation; LOOCV = leave-one-out cross validation; LSTM = long short-term memory; DCGAN = deep convolutional generative adversarial network; DC-AL GAN = DCGAN with AlexNet; VGG = Visual Geometry Group (algorithm).

Statistical measures: AUC = area under the receiver operator characteristic curve; BA = balanced accuracy; CI = confidence intervals.

*some data appears mathematically discrepant within publication

**unclear or discrepant information (e.g. time after chemoradiotherapy)

**Table 2.** Studies applying machine learning models to baseline images (or genomic alterations) to serve as prognostic biomarkers to predict subsequent treatment response

| Author(s) | Prediction | Reference | Dataset | Method | Features selected | Test set performance |
|---|---|---|---|---|---|---|
| Wang S et al., 2019[39] | True progression vs. Pseudoprogression (immunotherapy for EGFRvIII mutation) Baseline prediction | histopathology | model testing set = 10 DTI, DSC & 3D-EPSI | Prospective. 1 centre Multivariate logistic regression | CL, CBV, FA parameters | Insufficient published data to determine diagnostic performance (per lesion results available recall = 0.86 specificity = 0.60) |
| Yang K et al., 2019[40] | True progression vs. not (stable disease, partial & complete response & pseudoprogression) Baseline prediction | Imaging follow-up | training = 49 Genomic alterations | 1 centre Analysis including Gene Set Enrichment Analysis (GSEA) | Genomic alterations including CDKN2A & EGFR mutations | No test set (Insufficient published data to determine diagnostic performance from training dataset. 1-year PFS for responder 45% & non-responder 0%) |
| Lundemann M et al., 2019[41] | Early recurrence vs. not (voxel-wise) Baseline prediction | Mixture of histopathology & imaging follow-up | training = 10 18F-FET PET/CT 18F-FDG PET/MRI $T_1$C, $T_2$/FLAIR, DTI, DCE | Prospective. 1 centre Multivariate logistic regression LOOCV | FET, FDG, MD, FA, F, $V_b$, $V_e$, $K_i$, & MTT parameters | No test set (Insufficient published data to determine diagnostic performance from training dataset. Voxel-wise recurrence probability AUC 0.77) |

EGFR = Epidermal growth factor receptor; EGFRvIII = EGFR variant III; CDKN2A = cyclin-dependent kinase Inhibitor 2A.

Magnetic resonance imaging sequences: $T_1$C = post contrast $T_1$-weighted; $T_2$ = $T_2$-weighted; FLAIR = fluid-attenuated inversion recovery; DSC = dynamic susceptibility-weighted; DTI = diffusor tensor imaging; DCE = dynamic contrast-enhanced.
Additional imaging techniques: 3D-EPSI = 3D echo planar spectroscopic imaging; 18F-FET = [18F]-fluoroethyl-L-tyrosine; 18F-FDG = [18F]-fluorodeoxyglucose; PET/CT = positron emission tomography and computed tomography; PET/MRI = positron emission tomography and magnetic resonance imaging.
Magnetic resonance imaging parameters: CBV = cerebral blood volume; FA = fractional anisotropy; MD = mean diffusivity; F = blood flow; $V_b$ = vascular blood volume; $V_e$ = extra-vascular, extra-cellular blood volume; $K_i$ = vascular permeability; MTT = mean transit time; CL = linear anisotropy.

LOOCV = leave-one-out cross validation.
AUC = area under the receiver operator characteristic curve.
PFS = progression free survival.

### 3.2 Treatment response

Table 1 shows a variety of treatment response target conditions that individual studies were designed to predict. Approximately a quarter of studies assigned only the first 12 weeks after treatment as the time interval when pseudoprogression occurs - the full 6-month interval when there might be pseudoprogression was not incorporated (4/15, 27%). Approximately a quarter of studies predicted PTRE as a target condition (4/15, 27%). No study predicted radiation necrosis alone. Some studies analysed high grade glioma whereas the majority analysed only glioblastoma (13/15, 87%).

### 3.3 Follow-up imaging and histopathology at re-operation

Most studies used a combination of follow-up imaging and histopathology at re-operation to determine whether there was progression or a mimic (10/15, 67%). Some used one reference standard for one decision (progression) and another for the alternative decision (mimic) – this and other idiosyncratic rules caused a high risk of bias in some studies in terms of the reference standard and patient selection.

### 3.4 Features

Most studies only analysed imaging features alone (13/15, 87%). Three studies used implicit feature engineering based on convolutional neural networks (3/15, 20%).

### 3.5 Test sets

One third of studies did not use hold-out test sets (5/15, 33%) with a high risk of bias for index test methodology. Four studies used external hold-out test sets (4/15, 27%). In these four studies the ranges of diagnostic accuracy measures were: recall = 0.70-1.00; specificity = 0.67-0.90; precision = 0.78-0.88; F1 score = 0.74-0.94; balanced accuracy = 0.74-0.83; AUC = 0.80-0.85.

### 3.6 Subgroup analysis: prognostic biomarkers to predict subsequent treatment response

Two studies were prospective and both had a small samples size (both n = 10). One study used genomic alterations alone as features to predict the MRI treatment response. Diagnostic accuracy measures could not be determined because of study design. The unit of assessment in one study was per-lesion; another per-voxel; and another used a prognostic metric of 1-year progression free survival for the predicted treatment response groups. The studies are best considered as proof of concept.

## 4 Discussion

### 4.1 Summary of findings

There is only low level evidence [23] available to determine the diagnostic accuracy of glioblastoma treatment response monitoring biomarkers in adults developed through ML. The available evidence is at high risk of bias and there are concerns regarding applicability, particularly in determining treatment response status using the reference standards of histopathology at re-operation or follow-up imaging. There are similar and related concerns regarding study patient selection. One third of studies did not use hold-out test sets. Most studies used classic ML approaches with radiomic features; less than a quarter of studies used deep learning methodology.

### 4.2 Limitations

*4.2.1 Studies assessed*

The reference standards used led to a high risk of bias and concerns regarding applicability in all studies. Other than in the subgroup of prognostic biomarker studies, all studies were retrospective which increases the risk of confounding. Confounding factors, related to histopathological and imaging follow up reference standards, were second-line drug therapy and discontinuation of temozolomide– all of which were rarely accounted for. Similarly, steroid use was rarely accounted for and is a confounding factor for the imaging follow up reference standard. Some authors stated they followed RANO guidelines [4] which would overcome some of these limitations such as the use of steroids which is carefully integrated with the imaging assessment. However, a limitation in using RANO guidelines is that in some scenarios the upslope of any PTRE may be observed for second time over a short interval confounding assessment [21,22].

Patient selection was also problematic and is related to confounding – for example patients on second-line drug therapy should have been excluded from imaging follow-up response assessment.

One third of studies did not use hold-out test sets which are expected in ML studies for diagnostic accuracy assessment. Four studies, however, used external hold-out tests which is optimal practice.

*4.2.2 Review process*

Publication bias might have influenced the diagnostic accuracy of the monitoring biomarkers included in this review. Another limitation is that imaging reference standards such as RANO trial guidelines [4], and subsequent developments thereof [20], are themselves confounded and, when used, are rarely applied correctly [24]. In addition to the example described above regarding the upslope of PTRE [21,22], another limitation is not acknowledging that pseudoprogression occurs over a 6 month interval rather than a 3 month interval (although it is acknowledged that even 6 months is an arbitrary cut-off) [21]. For the purposes of the systematic review, study design would require imaging follow up of sufficient duration to take these issues into account. A limitation of this or other systematic reviews is therefore that it is highly challenging for studies to be designed with sufficient nuance to be at low risk of bias in relation to the reference standard. A further limitation of this review is that whilst histopathology at re-operation is not an entirely reliable reference standard [25], it was pragmatically chosen as acceptable in the absence of more reliable available reference standards at re-operation.

**4.3 Interpretation of the results in the context of other evidence**

There is good diagnostic performance of machine learning models that use MRI features to distinguish between progressive disease and mimics. As in the previous review, the diagnostic performance of ML using implicit features was not superior to ML using explicit features. However, the small numbers of patient included in studies, the high risk of bias and concerns of applicability in the study designs, and the low level of evidence given that the monitoring biomarker studies are retrospective, suggest that limited conclusions can be drawn from the data.

**4.4 Implications for future research and clinical practice**

The results show that glioblastoma treatment response monitoring biomarkers developed through ML are promising but are at an early phase of development and are not ready to be incorporated into clinical practice. All studies would benefit from improvements in the methodology highlighted above. Future studies would benefit from analytical validation using external hold-out tests as exemplified by several studies in the current review. Future studies would also benefit from larger datasets to reduce overfitting.

# 5 Conclusion

There are a range of ML-based solutions poised to become treatment response monitoring biomarkers for glioblastoma suitable for the clinic. To achieve this, the development and validation of ML models require large, well-annotated datasets where the potential for confounding in the study design has been carefully considered. Therefore, multidisciplinary efforts and multicentre collaborations are necessary [26].

# Acknowledgements

This work was supported by the Wellcome/EPSRC Centre for Medical Engineering [WT 203148/Z/16/Z], The Royal College of Radiologists and King's College Hospital Research and Innovation.

# 6 Appendix

**Supplementary Table S1.** MEDLINE, EMBASE and Cochrane Register search strategies. Recommendations for a sensitive search with low precision; with subject headings with exploded terms; and with no language restrictions, were followed [15].

---

MEDLINE (OVID). PubMed was included.

The search strategy for Title/Abstract terms used a combination of subject headings (MeSH terms) and keywords:

Database: Ovid MEDLINE(R) ALL <1946 to September 11, 2020>

Search Strategy:

--------------------------------------------------------------------------------

1   exp Glioblastoma/ (25451)

2   high grade glioma.mp. (2986)

3   pseudoprogression.mp. (633)

4   radiomics.mp. (2262)

5   exp Artificial Intelligence/ or exp Machine Learning/ or exp Neural Networks, Computer/ (99521)

6   exp Deep Learning/ (2761)

7   monitoring biomarker.mp. (71)

8   treatment response.mp. (29303)

9   imaging.mp. (2039219)

| | | |
|---|---|---|
| 10 | exp Magnetic Resonance Imaging/ or MRI.mp. (544944) | |
| 11 | pet.mp. (103253) | |
| 12 | exp Positron-Emission Tomography/ (61544) | |
| 13 | 9 or 10 or 11 or 12 (2117006) | |
| 14 | 1 or 2 or 3 (28462) | |
| 15 | 4 or 5 or 6 or 7 or 8 (130625) | |
| 16 | 13 and 14 and 15 (321) | |
| 17 | limit 16 to last 2 years (130) | |
| 18 | 13 and 14 (6464) | |
| **19** | **limit 18 to last 2 years (1241)** | |

***************************

strategy 17 was insensitive so strategy 19 was employed for final search

---

EMBASE (OVID).

Subject headings and keywords:

Database: Embase <1974 to 2020 Week 37>

Search Strategy:

--------------------------------------------------------------------------------

1  exp glioblastoma/ (68063)

2  high grade glioma.mp. (5411)

3  pseudoprogression.mp. (1225)

4  exp radiomics/ (1271)

5  exp machine learning/ or exp artificial intelligence/ (227658)

6  exp deep learning/ (9382)

7  monitoring biomarker.mp. (108)

8  exp treatment response/ (265476)

9  exp multiparametric magnetic resonance imaging/ or exp imaging/ or exp nuclear magnetic resonance imaging/ (1112723)

10  magnetic resonance imaging.mp. (925155)

| 11 | MRI.mp. (445714) |
| 12 | PET.mp. or exp positron emission tomography/ (249021) |
| 13 | 1 or 2 or 3 (72158) |
| 14 | 4 or 5 or 6 or 7 or 8 (492158) |
| 15 | 9 or 10 or 11 or 12 (1331815) |
| 16 | 13 and 15 (14315) |
| 17 | limit 16 to last 2 years (3209) |
| **18** | **limit 17 to exclude medline journals (479)** |

***************************

strategy 18 was employed for final search to prevent duplication from MEDLINE

Cochrane Register.

Epistemonikos review database included, protocols included, CENTRAL (Cochrane central register of controlled trials included which includes

https://www.ebscohost.com/nursing/products/cinahl-databases,

https://clinicaltrials.gov, https://www.who.int/ictrp/en/).

Subject headings and keywords:

Date Run:        13/09/2020 15:52:52

| ID | Search | Hits |
|---|---|---|
| #1 | MeSH descriptor: [Glioblastoma] explode all trees | 628 |
| #2 | high grade glioma | 524 |
| #3 | pseudoprogression | 69 |
| #4 | imaging | 68926 |
| #5 | MeSH descriptor: [Magnetic Resonance Imaging] explode all trees | 7660 |
| #6 | MRI | 23753 |
| #7 | PET | 6912 |
| #8 | MeSH descriptor: [Positron-Emission Tomography] explode all trees | 988 |
| #9 | {OR #1-#3} | 1167 |
| #10 | {OR #4-#8} | 79474 |

| #11 {AND #9-#10} 297 |
|---|
| strategy 11 was employed for final search |
| Health Technology Assessment. https://database.inahta.org/<br><br>Subject headings and keywords:<br><br>ﾠ((("Glioblastoma"[mh]) OR (high grade glioma) OR (pseudoprogression)) |

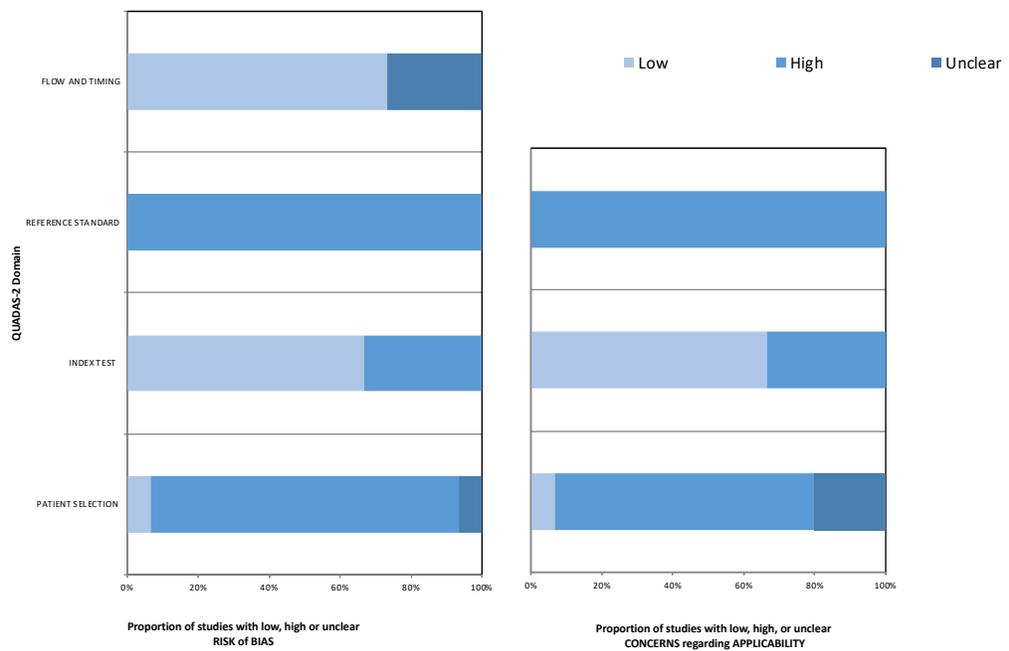

**Supplementary Figure S1.** Bar chart showing risk of bias and concerns of applicability assessment